\documentclass[prc,superscriptaddress,showpacs,psfig,showkeys,nofootinbib]{revtex4}
\usepackage{amsmath}
\usepackage{amssymb}
\usepackage{amsfonts}
\usepackage{graphicx}
\usepackage{epsfig}
\usepackage{dcolumn}
\usepackage{hyperref}
\usepackage{epstopdf}
\usepackage{color}
\usepackage[utf8]{inputenc}
\usepackage{amsthm}
\usepackage{fontenc}
\usepackage{bm}
\RequirePackage{slashed}\usepackage{cancel}
\usepackage[normalem]{ulem}

\usepackage{array,booktabs}
\newcolumntype{C}{>{$\displaystyle}c<{$}}
\usepackage{soul}
\usepackage{color,soul}
\newcommand{\red}{\color{red}}

\usepackage{ulem}

\begin{document}


\title{Slicing Pomerons in ultraperipheral collisions using forward neutrons from nuclear breakup}

\author{M. Alvioli}
\affiliation{Consiglio Nazionale delle Ricerche,
Istituto di Ricerca per la Protezione Idrogeologica,
via Madonna Alta 126, I-06128, Perugia, Italy}
\affiliation{
Istituto Nazionale di Fisica Nucleare, Sezione di Perugia,
via Pascoli 23c, I-06123, Perugia, Italy}

\author{V. Guzey}
\affiliation{University of Jyvaskyla, Department of Physics, P.O. Box 35, FI-40014 University
of Jyvaskyla, Finland and Helsinki Institute of Physics, P.O. Box 64, FI-00014 University of Helsinki, Finland}

\author{M. Strikman}
\affiliation{Pennsylvania  State  University,  University  Park,  PA,  16802,  USA}

\date{\today}

\begin{abstract}

We argue that measurements of forward neutrons from nuclear breakup in inclusive high energy photon-nucleus ($\gamma A$) scattering provide a novel 
way to study small-$x$ dynamics of QCD in heavy-ion ultraperipheral collisions (UPCs). 
Using models for hadronic fluctuations of the real photon and neutron emission in nuclear fragmentation, 
we calculate the distribution over the number of evaporation neutrons produced in $\gamma Pb$ collisions at the LHC. We 
show that it is correlated with 
the number of wounded nucleons (inelastic collisions) and, hence, can
constrain the mechanism of nuclear shadowing and its $x$ dependence.

\end{abstract}

\maketitle

\section{Introduction and motivation}
\label{sec:intro}

Understanding of the QCD dynamics of hard high energy interactions and the structure of nuclei and nucleons
is one of the main directions of theoretical and experimental studies at the Large Hadron Collider (LHC) and
the Relativistic Heavy Ion Collider (RHIC). Of particular interest is the limit of very small momentum
fractions $x$, when the linear Dokshitzer-Gribov-Lipatov-Altarelli-Parisi (DGLAP)  approximation is expected
to break down~\cite{Dokshitzer:1978hw,Gribov:1983ivg} and a regime close to the black disk limit (BDL)~\cite{Frankfurt:2001nt} 
may set in.
Its observation is one of the prime  objectives of the
planned Electron-Ion Collider (EIC) at Brookhaven National Laboratory~\cite{Accardi:2012qut,AbdulKhalek:2021gbh},
which will ultimately reach $x$ $\sim$ 10$^{-3}$ for momentum transfers of a few GeV. At the same time, it was pointed out some time ago that
ultraperipheral collisions (UPCs) of two ions  at the LHC, where a photon emitted by
one of the nuclei interacts with the other nucleus, allow one to probe  down to $x \sim 10^{-5} - 10^{-4}$,
depending on a particular reaction channel and the detector geometry~\cite{Strikman:2005yv,Baltz:2007kq}.

Electron-nucleus collisions at the EIC and UPCs of heavy ions at the LHC present 
two options for studying small-$x$ dynamics, which are largely complementary. At the LHC practically all data are collected for
one heavy nucleus and one cannot directly access the dependence of cross sections on the virtuality of the probe. 
At the same time, one can reach very small $x$, which is provided by a wide rapidity coverage of the LHC detectors
and the large invariant photon-nucleus collision energies exceeding by a factor of 100 the design energies
at the EIC.

Over the last decade the data taken in the LHC and RHIC kinematics discovered a significant nuclear suppression of coherent
$J/\psi$ photoproduction in Pb-Pb and Au-Au UPCs  compared to the impulse approximation
prediction~\cite{Guzey:2024gff}.
%\cite{ALICE:2013wjo,ALICE:2012yye,CMS:2016itn,ALICE:2021gpt,ALICE:2019tqa,LHCb:2022ahs,LHCb:2021bfl,CMS:2023snh,ALICE:2023jgu,STAR:2023gpk}.
When interpreted in the leading twist approximation (LTA)~\cite{Frankfurt:2011cs}, it amounts to strong
gluon nuclear shadowing~\cite{Guzey:2013xba,Guzey:2013qza}:
\begin{equation}
  R_A^g(x,Q^2)= \frac{g_A(x,Q^2)}{Ag_N(x,Q^2)} < 1 \,,
  \label{eq:R_A}
\end{equation}
where $g_A(x,Q^2)$ and $g_N(x,Q^2)$ are the nucleus and nucleon gluon densities, respectively. Typical numbers reported
by the LHC experiments, see~\cite{Guzey:2024gff} for references,
%~\cite{ALICE:2013wjo,ALICE:2012yye,CMS:2016itn,ALICE:2021gpt,ALICE:2019tqa,LHCb:2022ahs,LHCb:2021bfl,CMS:2023snh,ALICE:2023jgu}
correspond to  
\begin{eqnarray}
  R^g_{\rm Pb}(x=10^{-3}, Q^2_{\rm eff}=3\, {\rm GeV}^2) & \approx & 0.6 \,, \nonumber\\
  R^{g}_{\rm Pb}(x=10^{-4}, Q^2_{\rm eff}=3\, {\rm GeV}^2) &\approx & 0.5 \,,
  \label{eq:R_g}
\end{eqnarray}
with a similar suppression extending down to $x$ $\sim$ 10$^{-5}$. Here $Q_{\rm eff}$ is the effective resolution scale determined by the charm quark mass. 
These values of $R^{g}_{\rm Pb}$ agree very well with the LTA
predictions for nuclear shadowing made more than 10 years ago~\cite{Frankfurt:2011cs}.
Note that this interpretation of the $J/\psi$ UPC data is complicated at the next-to-leading order (NLO) of the perturbative
expansion in powers of $\log Q^2$ (perturbative QCD) due to large cancellations between the leading-order (LO) and NLO gluon
terms and a numerically important quark contribution~\cite{Eskola:2022vpi,Eskola:2022vaf}. A way to stabilize the
perturbation series and restore the gluon dominance in this process on the proton target was suggested in~\cite{Jones:2015nna,Jones:2016ldq}.

Other hard UPC processes considered in the literature are inclusive~\cite{Strikman:2005yv,Guzey:2018dlm} and
diffractive~\cite{Guzey:2016tek} dijet photoproduction, which are still to produce final results~\cite{ATLAS:2017kwa,ATLAS:2022cbd},
timelike Compton scattering~\cite{Pire:2008ea,Schafer:2010ud,Peccini:2021rbt}, and heavy quark photoproduction~\cite{Klein:2002wm,Goncalves:2009ey,Goncalves:2017zdx}.
Overall these findings confirm the conclusion of~\cite{Baltz:2007kq} that UPCs provide a very effective tool to access the
small-$x$ dynamics of the strong  interactions and the nuclear structure  in hard, semi-hard and soft regimes of QCD.

So far the experimental studies of UPCs have mainly been focusing on coherent and incoherent production of light and heavy
vector mesons.
In this paper, we would like to outline several possible directions of future UPC studies, which were not discussed in
the review~\cite{Baltz:2007kq} due to the lack of experimental confirmations of large nuclear shadowing. We explore for
the first time possibilities of testing the small-$x$ nuclear shadowing dynamics by measuring the rates of forward neutron
production from nuclear breakup in the zero degree calorimeters (ZDCs) at the LHC. Our numerical studies demonstrate that
the number of produced neutrons is correlated with the number of wounded nucleons (inelastic photon-nucleon interactions), which 
presents a complementary way to study the mechanism of nuclear shadowing.
In particular, as follows from the title of this paper, it allows one 
to control the number of unitarity cuts of diffractive exchanges (Pomerons), which build up the 
effect of nuclear shadowing. 

This paper is organized as follows. In Sec.~\ref{sec:AGK} we briefly summarize expectations based on applications of the
Abramovski-Gribov-Kancheli (AGK) theorem to photon-nucleus scattering, model the distribution over the number of wounded nucleons
and estimate its average value in the current UPC kinematics. Section~\ref{sec:ZDC} presents our predictions for
the distributions over the number of emitted forward neutrons from nuclear breakup in inelastic photon-nucleus scattering. 
Our conclusions and outlook are given Sec.~\ref{sec:conclusions}.

\section{Abramovski-Gribov-Kancheli (AGK) cutting rules, nuclear shadowing and the number of wounded nucleons in $\gamma A$ scattering}
\label{sec:AGK}

Part of the results discussed in this section have been presented in~\cite{Alvioli:2016gfo}; we summarize them below for completeness since they are needed for new results in Sec.~\ref{sec:ZDC}.

It was demonstrated by Abramovski, Gribov and Kancheli in 1973~\cite{Abramovsky:1973fm} that different unitary cuts
of the diagrams corresponding to multi-Pomeron (color singlet) exchanges result in different multiplicities of produced
particles in the central rapidity region and that the absorptive part of the amplitude can be expressed in terms of a
small number of cut diagrams. 
They are related by combinatorial factors, which is known as
the Abramovski-Gribov-Kancheli (AGK) cutting rules or the AGK cancellation. For the interpretation of the AGK rules in QCD and other
effective field theories, see Refs.~\cite{Treleani:1994at,Jalilian-Marian:2004vhw,Gelis:2006yv,Kovner:2006wr,Nikolaev:2006mx}.

The application of the AGK cutting rules to photon-nucleus scattering allows one to express the nuclear
shadowing correction to the total nuclear cross section $\sigma_{\rm tot}^{\gamma A}$ in terms of the diffractive cross
section on individual nucleons~\cite{Frankfurt:2011cs,Frankfurt:1998ym}. 

Another important application of the AGK cutting rules involves the total hadron-nucleus inelastic cross section defined as the difference between the total and total elastic (coherent plus incoherent) cross sections, which is obtained using the unitary form of the Glauber theory~\cite{Bertocchi:1976bq}. Generalizing this result to photon-nucleus scattering by taking into account that the photon takes part in the strong interactions by means of its hadronic fluctuations, 
the photon-nucleus total inelastic cross section can be presented in the following form~\cite{Alvioli:2016gfo}
\begin{equation}
  \sigma_{\rm inel}^{\gamma A}=\sum_{\nu=1}^A \sigma_{\nu} \,,
  \label{eq:sigma_A}
\end{equation}
where 
\begin{equation}
  \sigma_{\nu}=\frac{A!}{(A-\nu)! \nu!} \int d^2 \vec{b} \int d\sigma
  P_{\gamma}(\sigma) (\sigma_{\rm inel} T_A(\vec{b}))^{\nu} (1-\sigma_{\rm inel} T_A(\vec{b}))^{A-\nu} \,.
  \label{eq:sigma_n}
\end{equation}
In Eq.~(\ref{eq:sigma_n}), $\vec{b}$ is the impact parameter (transverse coordinate) of the interacting
nucleon, $T_A(\vec{b})=\int dz \rho_A(\vec{b},z)$, where $\rho_A(\vec{b},z)$ is the nuclear density normalized to unity,
and $\sigma_{\rm inel}=0.85 \,\sigma$ is the inelastic cross section for the interaction of a hadronic fluctuation of the photon with a target nucleon, which is based on the estimate that the $\rho$ meson-nucleon elastic cross section 
constitutes approximately 15\% of the total one.
The cross sections $\sigma_{\nu}$ in Eqs.~(\ref{eq:sigma_A}) and (\ref{eq:sigma_n}) 
represent the physical process,
where $\nu$ nucleons undergo inelastic scattering, while the remaining $A-\nu$ nucleons provide absorption. 
In the literature, one uses the term ``wounded nucleons''~\cite{Bialas:1976ed} and the notation
$\nu=N_{\rm coll}$.

The distribution $P_{\gamma}(\sigma)$ gives the probability density for hadronic fluctuations of the real
photon to interact with nucleons with the cross section $\sigma$~\cite{Alvioli:2016gfo,Frankfurt:2022jns}.
While the shape of $P_{\gamma}(\sigma)$ cannot be calculated from the first principles, one can reliably model it using 
constraints on its first moments and the small-$\sigma$ and large-$\sigma$ limits. Indeed, the total photon-proton
cross section $\sigma_{\gamma p}(W)$ and the cross section of photon diffractive dissociation on the proton
$d\sigma_{\gamma p \to X p}(W,t=0)/dt$ constrain the first two moments of $P_{\gamma}(\sigma)$ as follows
\begin{eqnarray}
\sigma_{\gamma p}(W) &=& \int d\sigma  P_{\gamma}(\sigma) \sigma \,, \nonumber\\
\frac{d \sigma_{\gamma p}(W,t=0)}{dt} &=& \frac{1}{16 \pi} \int d\sigma  P_{\gamma}(\sigma) \sigma^2 \,,
\label{ea:Psigma1}
\end{eqnarray}
where $W$ is the invariant photon-nucleon center-of-mass-energy. Further, in the small-$\sigma$ limit, one can 
express $P_{\gamma}(\sigma)$ in terms of the quark-antiquark component of the photon light-cone wave function and the color dipole cross section, which results in $P_{\gamma}(\sigma) \propto 1/\sigma$. In the opposite limit of large $\sigma$,
the photon behaves as a superposition of the $\rho$, $\omega$ and $\phi$ vector mesons in the spirit of the vector meson dominance model and, hence, $P_{\gamma}(\sigma)$ can be modeled using hadronic (cross section) fluctuations in $\rho$ mesons, which in turn are related to those for pions. Finally, the small-$\sigma$ and large-$\sigma$ regimes can be smoothly interpolated. Note that this matching is achieved best, when 
the light quark masses $m_q$ are taken to be those of the constituent quarks, $m_q \sim 300$ MeV.
For details, see~\cite{Alvioli:2016gfo,Frankfurt:2022jns}.

\begin{figure}[t]
  \centerline{%
    \includegraphics[width=9cm]{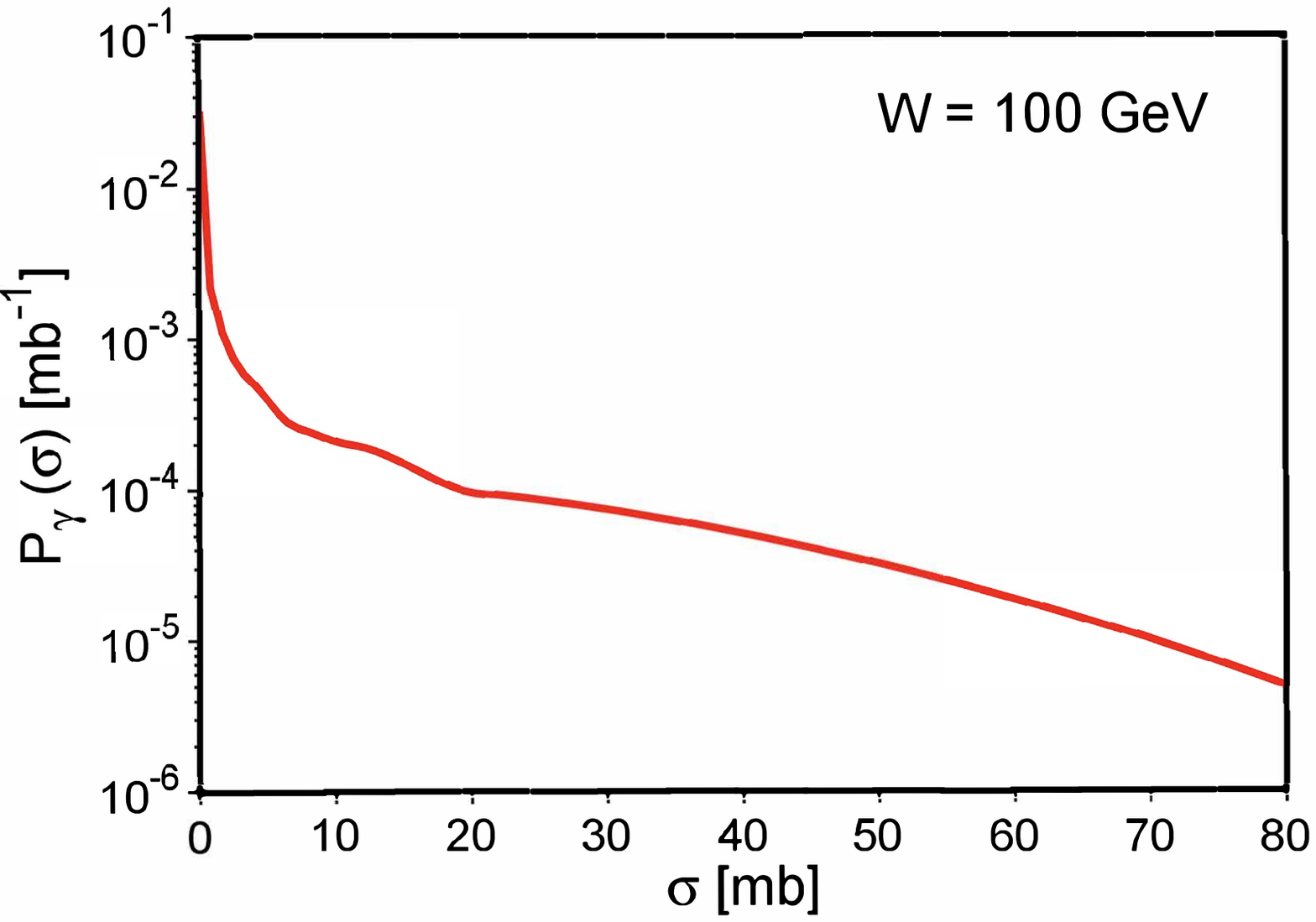}
    \includegraphics[width=9cm]{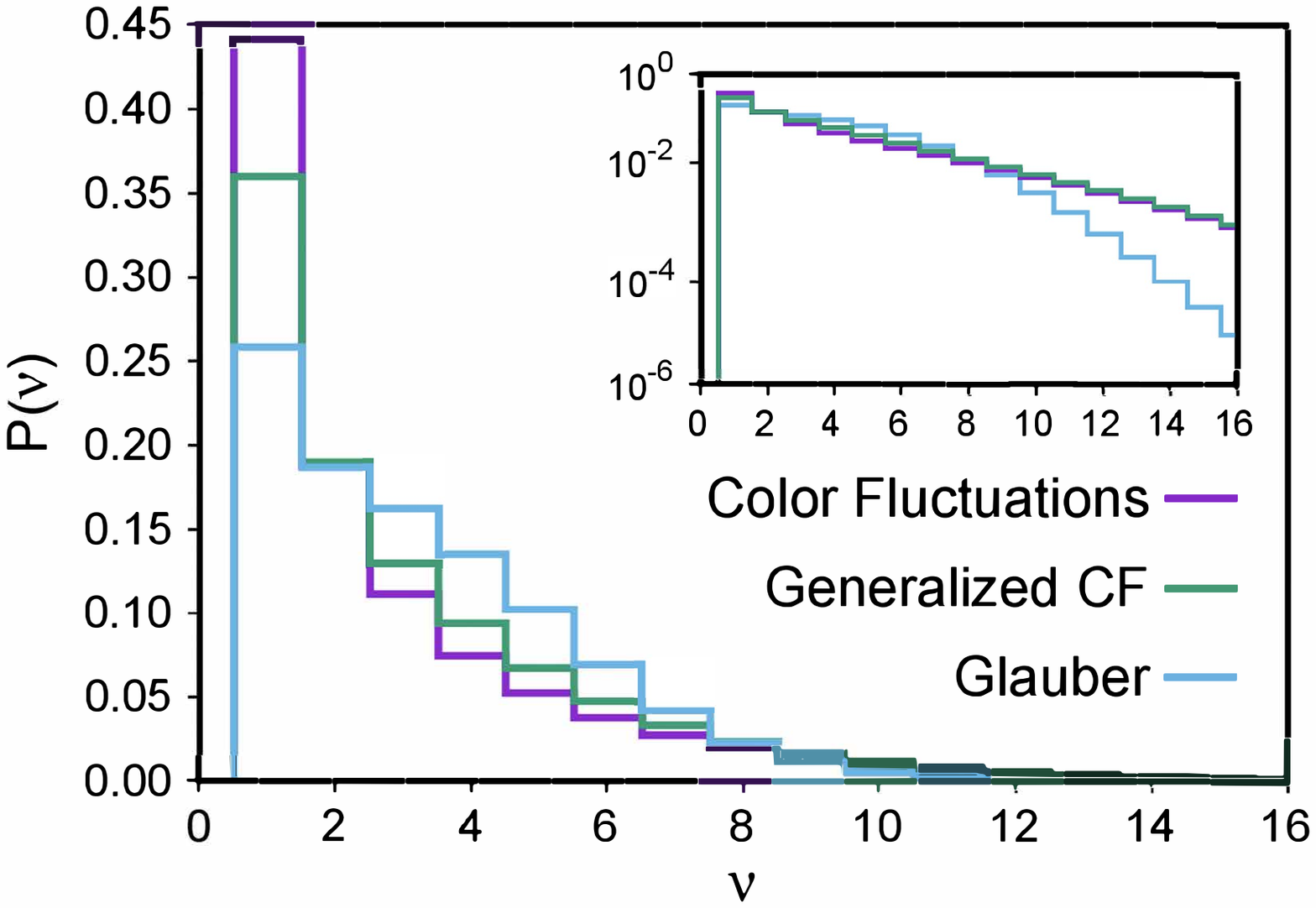}}
  \caption{(Left) The probability density $P_{\gamma}(\sigma)$ for hadronic fluctuations of the real photon to interact with nucleons with the 
  cross section $\sigma$ at $W=100$ GeV.
  (Right) The $P(\nu)$ distribution over the number of wounded nucleons
  (inelastic interactions) $\nu$ in inelastic photon-nucleus (Pb) scattering. The three curves correspond
  to the three models for $P_{\gamma}(\sigma)$, 
  see text for details. The insert emphasizes the region of large $\nu$.
  The figures are adopted from~\cite{Alvioli:2016gfo}.}
\label{fig:Pnu}
\end{figure}

The left panel of Fig.~\ref{fig:Pnu} presents the resulting distribution $P_{\gamma}(\sigma)$ as a function of $\sigma$ at $W=100$ GeV. Its 
shape and normalization are constrained by the procedure outlined above, the parametrization of $\sigma_{\gamma p}(W)$~\cite{ParticleDataGroup:2014cgo},
and the data on $d\sigma_{\gamma p \to X p}(W,t=0)/dt$~\cite{Chapin:1985mf}. Since the $W$ dependence of $P_{\gamma}(\sigma)$ is weak, the presented distribution
is applicable to a wide range of energies probed in heavy-ion UPCs at the LHC. 
Note that the distribution $P_{\gamma}(\sigma)$ parametrizes the so-called resolved photon contribution to photon-induced scattering and does contain
the direct photon contribution.

Note that the notion of cross section (color) fluctuations in hadron-nucleus scattering has found important phenomenological applications 
and confirmation in jet production in proton-nucleus scattering at the LHC and deuteron-nucleus scattering at RHIC~\cite{Alvioli:2014eda,Alvioli:2017wou}
as well as in pion and photon production in $d+{\rm Au}$ scattering at RHIC~\cite{Perepelitsa:2024eik}.

Using Eqs.~(\ref{eq:sigma_A}) and (\ref{eq:sigma_n}), one can readily define the $P(\nu)$ probability distribution for the number of wounded
nucleons $\nu$ in inelastic photon-nucleus scattering as follows~\cite{Alvioli:2016gfo},
\begin{equation}
  P(\nu)=\frac{\sigma_{\nu}}{\sum_{\nu=1}^A \sigma_{\nu}} \,,
  \label{eq:Pnu}
\end{equation}
where $\sigma_{\nu}$ is given by Eq.~(\ref{eq:sigma_n}). To calculate it, we use the Monte Carlo generator for nucleon configurations in complex nuclei~\cite{Alvioli:2009ab}, which also includes 
nucleon-nucleon correlations in the nucleus wave function~\cite{Alvioli:2008rw,Alvioli:2007zz}, and the
Gribov-Glauber model for photon-nucleus scattering, where the hadronic structure of the photon is described by 
$P_{\gamma}(\sigma)$.
The resulting distribution $P(\nu)$ as a function of $\nu$ for lead (Pb) is shown by the curve labeled ``Color Fluctuations'' in
the right panel of Fig.~\ref{fig:Pnu}.

Our modeling of small-$\sigma$ hadronic fluctuations of the photon is based on the quark-antiquark component of the photon wave function and corresponds to a weak nuclear shadowing, which disagrees with the observed strong gluon nuclear shadowing, see Eq.~(\ref{eq:R_g}). To take it into account,
we model the nuclear suppression of the dipoles with $\sigma \leq \sigma_0=20$ mb by the factor of $R^g_A$, which leads to the modified 
distribution $\tilde{P}_{\gamma}(\sigma)$,
\begin{equation}
\tilde{P}_{\gamma}(\sigma)=\left[R^g_A(x,Q^2_{\rm eff})\theta(\sigma_0-\sigma) +\theta(\sigma-\sigma_0)\right] P_{\gamma}(\sigma) \,,
\label{eq:P_gamma_mod}
\end{equation}
where $x=Q^2_{\rm eff}/W^2$ and $P_{\gamma}(\sigma)$ is shown in Fig.~\ref{fig:Pnu} (left panel).
The distribution $P(\nu)$ corresponding to $\sigma_{\nu}$ calculated using $\tilde{P}_{\gamma}(\sigma)$ is given by the curve ``Generalized CF" in the right panel of Fig.~\ref{fig:Pnu}. 

Finally, to test the importance of cross section (color) fluctuations in the real photon, we calculated $P(\nu)$ neglecting these
fluctuations and using 
\begin{equation}
P_{\gamma}(\sigma)=\delta(\sigma-25\, {\rm mb}) 
\label{eq:P_gamma_mod2}
\end{equation}
in Eq.~(\ref{eq:sigma_n}). The result is given by the curve ``Glauber'' in Fig.~\ref{fig:Pnu}.
One can see from this figure that color fluctuations significantly increase the distribution $P(\nu)$ at small and especially large $\sigma$; the latter is emphasized in the insert.

In the total inelastic photon-nucleus cross section, the AGK cancellations manifest themselves as the observation that the average number of wounded nucleons $\langle N_{\rm coll} \rangle$
is inversely proportional to the nuclear shadowing factor. Generalizing the result of \cite{Bertocchi:1976bq} for hadron-nucleus 
scattering to the case of photon-induced scattering, one obtains~\cite{Alvioli:2016gfo}
\begin{equation}
  \langle N_{\rm coll} \rangle \equiv \sum_{\nu=1}^A P(\nu) \nu =\frac{\sum_{\nu=1}^A \nu \sigma_{\nu}}{\sum_{\nu=1}^A  \sigma_{\nu}} = \frac{A \sigma_{\rm inel}^{\gamma N}}{\sigma_{\rm inel}^{\gamma A}} \,,
  \label{eq:n_av}
\end{equation}
where $\sigma_{\rm inel}^{\gamma N}$ is the photon-nucleon inelastic cross section. Considering a particular
hard process in inelastic photon-nucleus scattering that probes the nuclear gluon distribution,
e.g., inclusive charmonium (bottomonium) production $\gamma+A \to J/\psi (\Upsilon)+X$ or
inclusive heavy-quark dijet production $\gamma+A \to Q {\bar Q}+X$, one obtains using Eq.~(\ref{eq:n_av}){\red:}
\begin{equation}
  \langle N_{\rm coll} \rangle \approx \frac{1}{R^g_{\rm Pb}(x,Q^2)} \lesssim 2  \,.
  \label{eq:n_av_2}
\end{equation}
In this estimate we used the following considerations. In general, the effect of nuclear shadowing for the
nuclear inelastic cross section is somewhat larger than that for the total cross section. 
However, the theoretical uncertainties of the leading twist approximation~\cite{Frankfurt:2011cs}
largely mask the differences and make the shadowing effects approximately equal (within
uncertainties) in the two cases. Finally, in the last step, we used the results of Eq.~(\ref{eq:R_g}). 

\begin{figure}[t]
  \centerline{%
    \includegraphics[width=10cm]{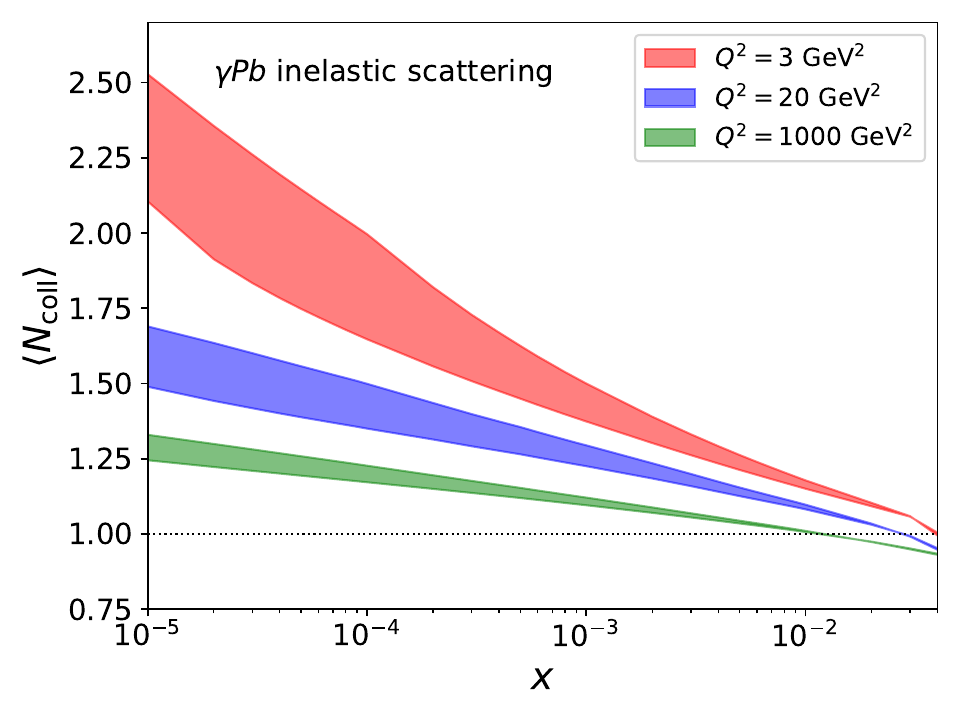}}
  \caption{The LTA predictions for the average number of wounded nucleons 
$\langle N_{\rm coll} \rangle$ in inelastic photon-nucleus (Pb) scattering as a function of $x$ at $Q^2=3$, 
20, and 1000 GeV$^2$. }
  \label{fig:Ncoll_x_final}
\end{figure}

Figure~\ref{fig:Ncoll_x_final} shows predictions of the leading twist approximation (LTA) for the average number of wounded nucleons 
$\langle N_{\rm coll} \rangle=1/R^g_{\rm Pb}(x, Q^2)$ in inelastic photon-nucleus (Pb) scattering as a function of $x$ at $Q^2=3$, 
20, and 1000 GeV$^2$. These values of $Q^2$ correspond to photoproduction $J/\psi$, $\Upsilon$, and high-$p_T$ dijets, respectively.
One can see from the figure that in the discussed kinematics, the average number of wounded nucleons is modest. As a result, the series
in Eq.~(\ref{eq:n_av}) converges rather rapidly. In particular, we have checked that it is saturated by first six terms with a 5\% precision. 
Note, however, that the convergence slows down in the limit of small $x$.

Measurements of $\langle N_{\rm coll} \rangle$ present a new method to study nuclear shadowing in inelastic photon-nucleus 
scattering. Unlike the observables used so far, the constraint of Eq.~(\ref{eq:n_av_2}) indicates that one can perform a
``Pomeron surgery'' of nuclear shadowing by cutting a small number of Pomeron exchanges controlling the number of inelastic
interactions with target nucleons. As a result, it gives an opportunity for an experimental determination
of a small number of parameters quantifying nuclear shadowing, which leads to a systematic improvement of its theoretical
description. We elaborate on it in the following section.

\section{The distributions over the number of wounded nucleon and forward neutrons from nuclear breakup}
\label{sec:ZDC}

The average number of inelastic interactions $\langle N_{\rm coll} \rangle$ encodes information on the energy and scale
dependence of nuclear shadowing.
To obtain a more microscopic description of nuclear shadowing, 
one needs
to determine not only $\langle N_{\rm coll} \rangle$, but also the entire distribution over the number of wounded
nucleons. This can be done using experimental data on the neutron emission resulting
from nucleus fragmentation in a given UPC process, e.g., in inclusive quarkonium or dijet photoproduction in heavy-ion UPCs 
with an additional condition of $Xn$ neutrons in the zero degree calorimeter (ZDC) on the nuclear target side~\cite{ATLAS:2017kwa,ATLAS:2022cbd}.

Very little is known about the dynamics of neutron emission in high energy scattering off heavy nuclei. The ALICE
collaboration measured the distribution over the number of collisions $\langle N_{\rm coll}(E_T) \rangle$ in proton-nucleus scattering
as determined by the energy release ($E_T$) at central rapidities  and the neutron multiplicity  as a function of $E_T$~\cite{ALICE:2021poe}.  It was observed 
that  $\langle N_{\rm coll}(E_T) \rangle$ is linearly proportional to the number of evaporation neutrons $\langle M_n(E_T) \rangle$ for the same $E_T$ bins at least up to $\langle N_{\rm coll} \rangle \sim 10$.
Note that in our case, $\langle N_{\rm coll} \rangle$ is much lower, see Eq.~(\ref{eq:n_av_2}).

Another important observation made by the E665 experiment at Fermilab is that in muon-nucleus deep inelastic scattering
(DIS) in coincidence with detection of slow neutrons, $\mu^{-}+A \to n+X$, the average neutron multiplicity
$\langle M_n \rangle$ for the lead target is~\cite{E665:1995utr}
\begin{equation}
  \langle M_n \rangle \approx 5 \,.
  \label{eq:E665}
\end{equation}
This result has been understood in the framework of cascade models of nuclear DIS~\cite{Strikman:1998cc,Larionov:2018igy},
where soft neutrons are produced either directly in DIS on a bound nucleon or through a statistical decay (de-excitation)
of the excited residual nucleus, leading to neutron evaporation\footnote{The geometry of the heating is very different
from the case of $AA$ collisions, where in each collision a large portion of nucleons in each nucleus interacts and de-excitation
of the spectators only occurs close to the interaction surface~\cite{Alvioli:2010yk}.}. The latter mechanism depends strongly
on the hadron formation time: to describe the energy spectrum of emitted neutrons, one has to assume that only nucleons and pions with
momenta $\leq 1$ GeV/c could be involved in final-state interactions. A similar
conclusion was reached by M.~Baker (private communication, 2022) using the BeAGLE Monte Carlo generator~\cite{Chang:2022hkt}.

This suggests the following space-time picture of forward neutron production in high energy photon-nucleus scattering. 
Well before the target, the incoming photon fluctuates into long-lived hadronic components, which pass through the nucleus
and interact inelastically with several nucleons. It leads to the creation of holes in the nucleus
(particle-hole excitations in terminology of a nuclear shell model), which de-excite and cool the nucleus by evaporating
neutrons. It also produces a number of soft particles with the momenta less than 1 GeV/c, which in turn generate more
neutrons. 

The nucleon fragmentation weakly depends on the incident energy  due to Feynman scaling and, hence, it is not significantly affected
by the energy conservation constraint, which is important in the case, when one uses hadron multiplicities at central rapidities~\cite{ATLAS:2015hkr}. In
this case, the energy transferred to the rest of the nucleus,  which heats the residual nuclear system, is proportional
to $\langle N_{\rm coll} \rangle$. Since the Fermilab data~\cite{E665:1995utr} corresponds to the average momentum fraction $\langle x \rangle
= 0.015$, where the nuclear shadowing effect is small, one finds that $\langle N_{\rm coll} \rangle$ $\approx$ 1, see Eq.~(\ref{eq:n_av}). Thus,  every inelastic
photon-nucleon interaction results on average in 5 forward neutrons.

\begin{figure}[t]
  \centerline{%
    \includegraphics[width=9cm]{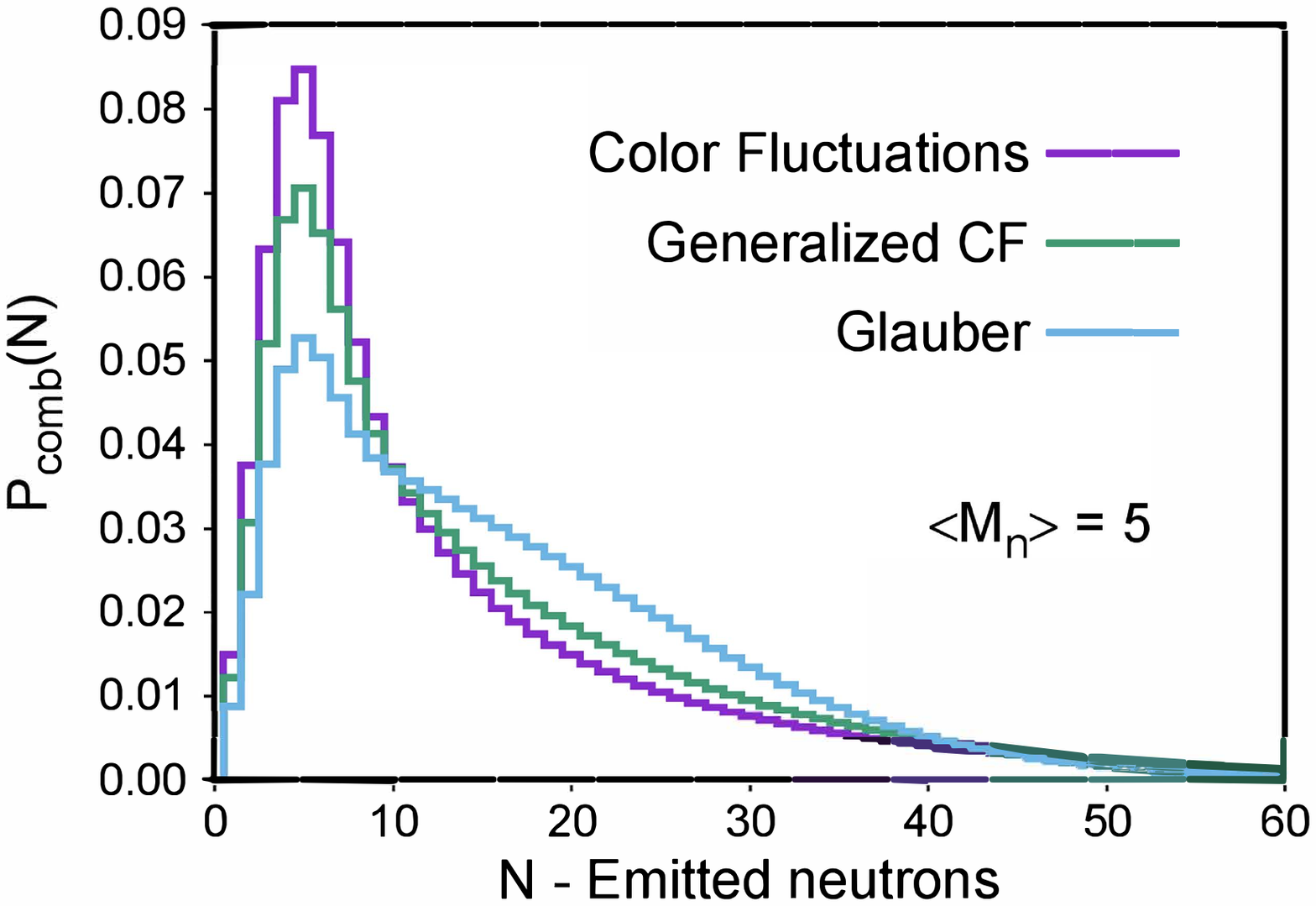}
    \includegraphics[width=9cm]{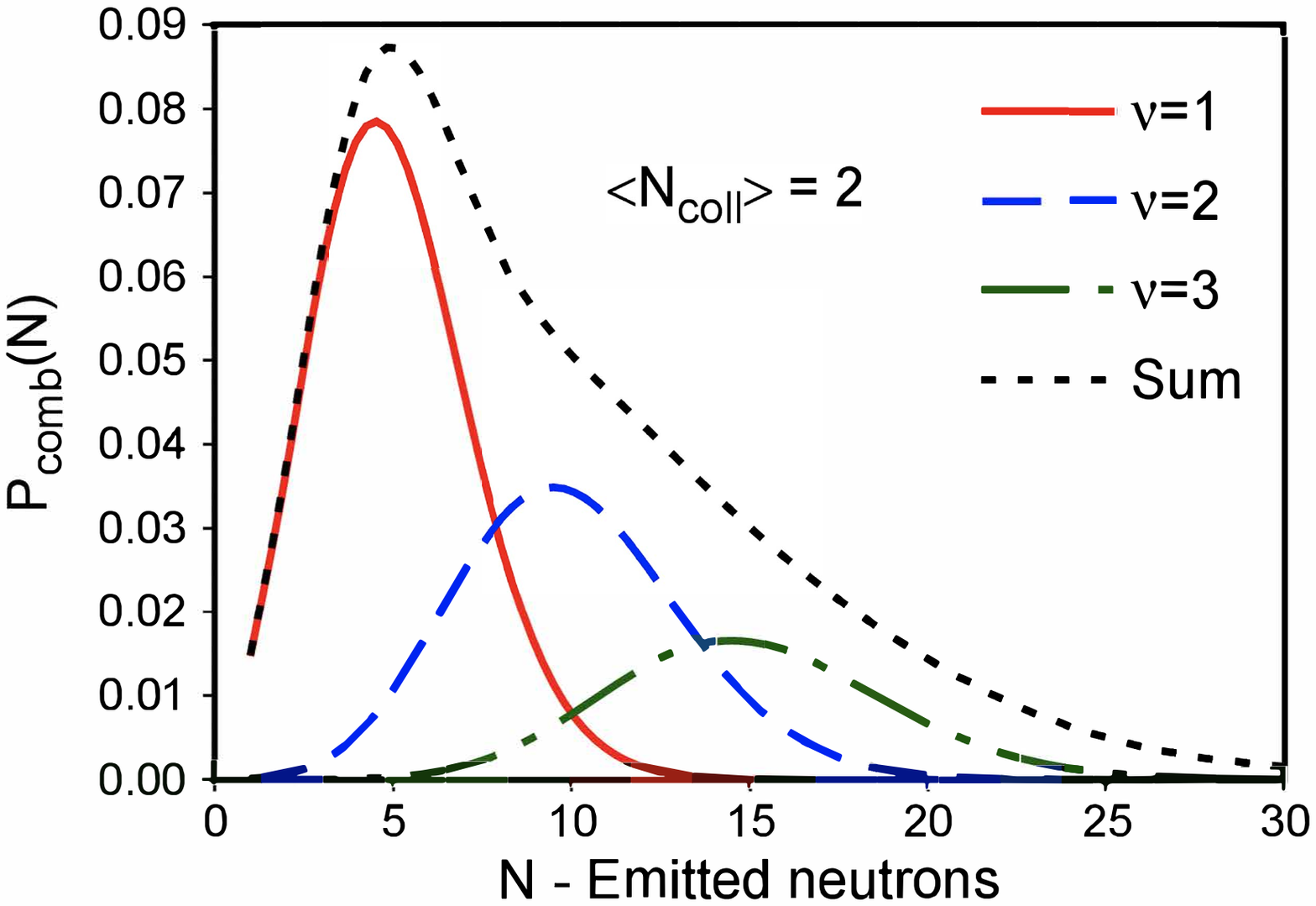}}
  \caption{
      (Left) The probability distribution of forward neutron emission $P_{\rm comb}(N)$~(\ref{eq:P_Mn}) 
      as a function of the number of emitted neutrons $N$ for $\langle M_n \rangle =5$. The three curves correspond to the three models for 
      $P_{\gamma}(\sigma)$ discussed in text.
      (Right) The contributions of $\nu=1,2,3$ wounded nucleons to $P_{\rm comb}(N)$ in the Glauber model for $P_{\gamma}(\sigma)$ 
      chosen to correspond to $\langle N_{\rm coll} \rangle =2$.  The black dotted curve labeled ``Sum'' gives the total $P_{\rm comb}(N)$.}
  \label{fig:frequency}
\end{figure}

We use this hypothesis and perform the following numerical test study. 
First, we consider a simple model, which assumes that the probability density of neutron emission is given by the Poisson distribution
and that each hole created in the target nucleus generates independently 
on average $\langle M_n \rangle$ neutrons. Therefore, the neutron probability distribution for $\nu = \langle N_{\rm coll} \rangle$ 
wounded nucleons is
\begin{equation}
P_{\rm Poisson}(N; \lambda=\nu \langle M_n \rangle)=\frac{(\nu \langle M_n \rangle)^{N}  e^{-\nu \langle M_n \rangle}}{N!} \,,
\label{eq:Poisson}
\end{equation}
where $N$ is the number of produced neutrons (neutron multiplicity). 

Then,
we combine the distribution over the number of wounded nucleons $P(\nu)$ discussed in Sec.~\ref{sec:AGK}
with the Poisson distribution of produced neutrons~(\ref{eq:Poisson}).
 The resulting probability
distribution of forward neutrons is given by the following expression,
\begin{equation}
P_{\rm comb}(N)=\sum_{\nu=1}^{A}P(\nu) P_{\rm Poisson}(N; \nu \langle M_n \rangle) \,.
\label{eq:P_Mn}
\end{equation}
The left panel of Fig.~\ref{fig:frequency} presents $P_{\rm comb}(N)$ as a function of forward neutrons 
$N$ for $\langle M_n \rangle =5$. The three curves correspond to the three models for the hadronic (color) fluctuations of the 
real photon, see the right panel of Fig.~\ref{fig:Pnu}. One can see from the figure that cross section fluctuations of the real photon
noticeably affect the shape of the neutron distribution: 
its maximum at small $N$ is more pronounced
compared to the ``Glauber'' result and is also
somewhat suppressed by the leading twist shadowing in the ``Generalized CF'' case.

Extraction of the contributions of individual $\nu$ (deconvolution) from the distribution $P_{\rm comb}(N)$ is feasible for not very large values of
$\sigma_{\rm inel}^{\gamma N}$. This is illustrated in the right panel of Fig.~\ref{fig:frequency}, which shows the contributions of $\nu=1,2,3$ wounded nucleons
to $P_{\rm comb}(N)$ in the case, when $P_{\gamma}=\delta(\sigma-\bar{\sigma})$ with $\bar{\sigma}$ chosen to correspond to $\langle N_{\rm coll} \rangle =2$, see
Fig.~\ref{fig:Ncoll_x_final}. The black dotted curve labeled ``Sum'' gives the total $P_{\rm comb}(N)$ in this model.
One can see from the figure that the peaks of these contributions are sufficiently separated, which gives a principal 
possibility to reconstruct the distribution over wounded nucleons $P(\nu)$. 
Then one can use an iterative procedure to
find individual $\sigma_{\nu}$. Indeed, assuming that the series in Eq.~(\ref{eq:sigma_A}) of dominates by the $\nu=1,2$ terms
(the limit of weak nuclear shadowing), one obtains 
\begin{equation}
\langle N_{\rm coll}-1 \rangle  \approx  \frac{\sigma_2}{\sigma_1} \approx \frac{A-1}{2} \frac{\langle \sigma_{\rm inel}^2 \rangle}{\langle \sigma_{\rm inel} \rangle}  
 \int d^2 \vec{b} \,T_A^2(\vec{b}) \,,
 \label{eq:sigma2}
\end{equation}
 where in the second equation, we used Eq.~(\ref{eq:sigma_n}). The ratio $\langle \sigma_{\rm inel}^2 \rangle/\langle \sigma_{\rm inel} \rangle$
 plays a central role in the leading twist approach to nuclear shadowing, where it determines its magnitude in the weak shadowing limit.
 Working along these lines, one can determine higher moments $\langle \sigma_{\rm inel}^n \rangle/\langle \sigma_{\rm inel} \rangle$, see~\cite{Alvioli:2014sba}, which built up 
 a full-fledged LTA shadowing correction. 
 
\begin{figure}[t]
  \centerline{%
    \includegraphics[width=9cm]{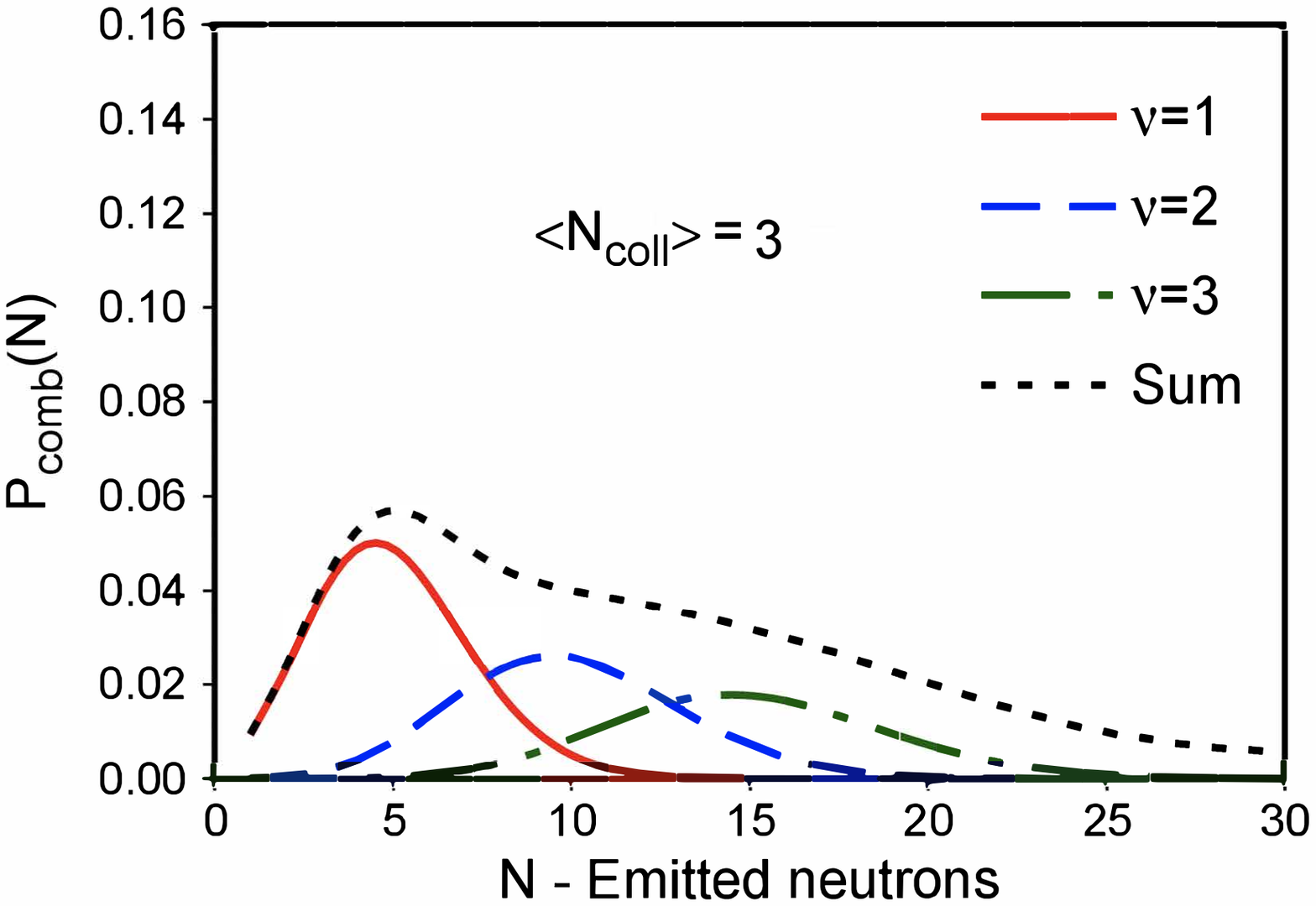}
    \includegraphics[width=9cm]{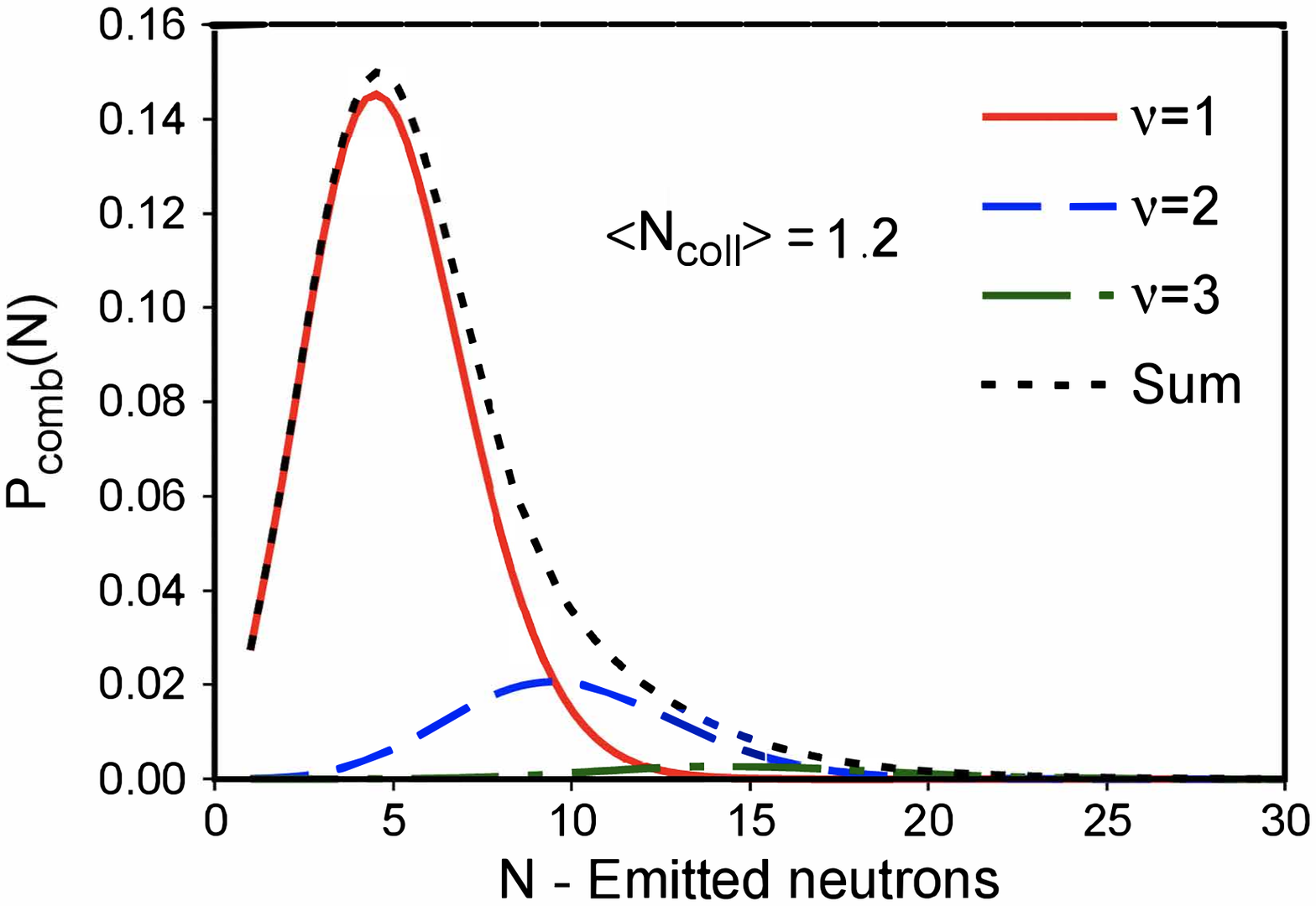}}
  \caption{
      The probability distribution of forward neutron emission $P_{\rm comb}(N)$ as a function of the number of emitted neutrons $N$ for $\langle M_n \rangle =5$.
      The left and right panels correspond to the strong nuclear shadowing with $\langle N_{\rm coll} \rangle=3$ and weak shadowing with $\langle N_{\rm coll} \rangle=1.2$, respectively. See Fig.~\ref{fig:frequency} for legend.}
  \label{fig:frequency2}
\end{figure}

The shape of the neutron distributions in Fig.~\ref{fig:frequency} strongly depends on the magnitude of nuclear shadowing, which in turn is correlated with $\langle N_{\rm coll} \rangle$ and the $x$-dependence
(energy dependence) of nuclear modification of nuclear PDFs in LTA: an increase of nuclear shadowing leads to the proportional increase of $\langle N_{\rm coll} \rangle$),
see Eq.~(\ref{eq:n_av_2}) and Fig.~\ref{fig:Ncoll_x_final},
and, hence, to a wider distribution $P_{\rm comb}(N)$ with important contributions of large $\nu$.
This is illustrated by the left panel of Fig.~\ref{fig:frequency2}, which shows 
$P_{\rm comb}(N)$ as a function of $N$ for $\langle N_{\rm coll} \rangle=3$. A comparison with the right panel of Fig.~\ref{fig:frequency}
demonstrates that $P_{\rm comb}(N)$ has become wider (the black dotted curve) because of an important contribution of $\nu \geq 2$ wounded nucleons.

Note that to reach a high accuracy in deconvolution of $P_{\rm comb}(N)$, one needs to calibrate the theoretical description against
the kinematics, where only one target nucleon is struck, e.g., using $\gamma+ A \to  {\rm 2\ jets} +X$ 
or quasi-elastic $J/\psi$ production for $x_A$ $\ge$ 0.01, where the effect of nuclear shadowing is small  and
$\langle N_{\rm coll} \rangle$ $\approx$ 1. It is supported by the results in the right panel of Fig.~\ref{fig:frequency2}, which show 
that $P_{\rm comb}(N)$ at $\langle N_{\rm coll} \rangle=1.2$ is dominated by the $\nu=1$ contribution.

In summary, an important qualitative effect predicted by our model is a strong increase of the multiplicity of neutrons detected in ZDCs with a decrease of $x$ from $x \sim 0.05$ corresponding to the right panel of Fig.~\ref{fig:frequency2}, where nuclear shadowing is small, to $x \sim  10^{-3}$ 
 corresponding to the left panel of Fig.~\ref{fig:frequency2}, where the shadowing effect is large.
After measurements of the neutron multiplicity for large $x$ are performed, it would be possible to test LTA predictions for the probabilities of
$1, 2, 3$ wounded nucleons as well as the assumption that emissions of neutrons generated by a removal of nucleons can be treated as independent.

\section{Conclusions and outlook}
\label{sec:conclusions}

In this paper, we suggest using measurements of forward neutrons from nuclear breakup in inclusive high energy photon-nucleus scattering in heavy-ion UPCs, e.g., 
charmonium (bottomonium) production $\gamma+A \to J/\psi (\Upsilon)+X$ or
heavy-quark dijet production $\gamma+A \to Q {\bar Q}+X$, as a novel way to study the QCD dynamics at small $x$. The key quantity is the number of inelastic photon-nucleon interactions (the number of wounded nucleons): its average value $\langle N_{\rm coll} \rangle$
is proportional to inverse of the gluon nuclear shadowing and its 
distribution is sensitive to details of nuclear shadowing.
Our numerical analysis suggests
that the number of forward neutrons from
nuclear breakup detected in the ZDC on the nuclear target side is rather unambiguously proportional to the number of wounded nucleons, 
which provides a practical opportunity for novel studies of nuclear shadowing and its $x$ dependence.

Using these processes, it would be possible to
explore effects related to proximity to the black disk limit of the strong interaction.
For example, one can study fragmentation of leading hadrons in $\gamma A$ scattering and look for suppression of their
multiplicity as a function of Feynman $x_F$ and $W$ as well as for
broadening of their transverse momentum distribution~\cite{Frankfurt:2001nt}.
These effects should be more pronounced for central collisions characterized by an enhanced activity in the ZDC. 
It should be possible
to construct from the data an analog of the central-to-peripheral $R_{CP}$ ratio of yields, which would probe the density dependence of fragmentation. 
It would also be useful to construct similar quantities for low-$p_T$ charm production.

Another interesting application is for multiparton interactions in proton-nucleus ($pA$) scattering. It was argued
in~\cite{Alvioli:2019kcy} that the single and double scattering can be separated using their dependence on the impact parameter:
the former is proportional to $A$, while the latter $\propto A^{4/3}$. However, since  both hard interactions are typically detected in a limited range
of rapidities $|y| \leq 3-4$,  centrality is difficult to determine from the transverse energy $E_T$ signal
because multiparton interactions also contribute to $E_T$.
 The use of forward neutrons in ZDCs would alleviate this problem.

One should point out that the neutrons detected in ZDCs can be a promising complementary way to determine centrality of various photon-nucleus and proton-nucleus inelastic collisions expanding the use of ZDCs beyond their current use in vector meson diffractive production
and for determining of centrality of the heavy-ion collisions. The main advantage of using forward neutrons rather than the transverse energy $E_T$ for the determination of centrality is a much larger distance in rapidity between the rapidity of the hard process and that of the process used for determination of the centrality.

One of the principal problems of using UPCs for studies of small-$x$ phenomena is a lack of the nucleon reference data at similar energies with the precision necessary to observe nuclear effects with a better than 10\% accuracy ($J/\psi$ exclusive photoproduction is a notable exception). Below we outline a possible strategy for overcoming this problem. Note that we are not aiming to optimize cuts or to account for the  energy resolution of ZDCs since this would require a dedicated Monte Carlo study.

One can separate events into two classes: peripheral events corresponding to $\langle N_{\rm coll} \rangle \le 2$  (we call it class ``L'') and more central events corresponding to $\langle N_{\rm coll} \rangle \ge 1.5-2$ and $\langle M_n \rangle  \sim 7 - 10$ (class ``H'').
If statistics is sufficient,  the lower limit for class ``H'' can be gradually increased, which will push up the average  number of wounded nucleons.
Then, the ratio of the number of events in the two classes, ${\hat R}=\rm Yield(H)/Yield(L)$, should quantify the effect of nuclear shadowing at small and large
impact parameters, which in principle probes the dependence of nuclear shadowing on the thickness of nuclear matter.
The promising channels for such an analysis include inclusive charm production with the transverse momentum in the range $p_T=5-20$ GeV/c and
production of soft particles with small $p_T < 0.5$ GeV/c. A comparison of the rates of these processes will allow one to study the transition between the soft and hard regimes and will serve as a consistency check of the description of small-$x$ dynamics in the
current models.

The methods presented in this paper can be readily generalized to the case of virtual photons and allow one to predict the distribution over the 
number of forward neutrons in inelastic photon-nucleus scattering at the EIC.

\acknowledgments
The research of V.G.~was funded by the Academy of Finland project 330448, the Center of Excellence in Quark Matter
of the Academy of Finland (projects 346325 and 346326), and the European Research Council project ERC-2018-ADG-835105
YoctoLHC. The research of M.S.~was supported by the US Department of Energy Office
of Science, Office of Nuclear Physics under Award No. DE- FG02-93ER40771.


\begin{thebibliography}{99}

%\cite{Dokshitzer:1978hw}
\bibitem{Dokshitzer:1978hw}
Y.~L.~Dokshitzer, D.~Diakonov and S.~I.~Troian,
%``Hard Processes in Quantum Chromodynamics,''
Phys. Rept. \textbf{58}, 269-395 (1980)
%doi:10.1016/0370-1573(80)90043-5
%918 citations counted in INSPIRE as of 14 Feb 2024

%\cite{Gribov:1983ivg}
\bibitem{Gribov:1983ivg}
L.~V.~Gribov, E.~M.~Levin and M.~G.~Ryskin,
%``Semihard Processes in QCD,''
Phys. Rept. \textbf{100}, 1-150 (1983)
%doi:10.1016/0370-1573(83)90022-4
%3101 citations counted in INSPIRE as of 18 Jan 2024

%\cite{Frankfurt:2001nt}
\bibitem{Frankfurt:2001nt}
L.~Frankfurt, V.~Guzey, M.~McDermott and M.~Strikman,
%``Revealing the black body regime of small x DIS through final state signals,''
Phys. Rev. Lett. \textbf{87}, 192301 (2001)
%doi:10.1103/PhysRevLett.87.192301
[arXiv:hep-ph/0104154 [hep-ph]].
%70 citations counted in INSPIRE as of 17 Jan 2024

%\cite{Accardi:2012qut}
\bibitem{Accardi:2012qut}
A.~Accardi, J.~L.~Albacete, M.~Anselmino, N.~Armesto, E.~C.~Aschenauer, A.~Bacchetta, D.~Boer, W.~K.~Brooks, T.~Burton and N.~B.~Chang, \textit{et al.}
%``Electron Ion Collider: The Next QCD Frontier: Understanding the glue that binds us all,''
Eur. Phys. J. A \textbf{52}, no.9, 268 (2016)
%doi:10.1140/epja/i2016-16268-9
[arXiv:1212.1701 [nucl-ex]].
%1548 citations counted in INSPIRE as of 18 Jan 2024

%\cite{AbdulKhalek:2021gbh}
\bibitem{AbdulKhalek:2021gbh}
R.~Abdul Khalek, A.~Accardi, J.~Adam, D.~Adamiak, W.~Akers, M.~Albaladejo, A.~Al-bataineh, M.~G.~Alexeev, F.~Ameli and P.~Antonioli, \textit{et al.}
%``Science Requirements and Detector Concepts for the Electron-Ion Collider: EIC Yellow Report,''
Nucl. Phys. A \textbf{1026}, 122447 (2022)
%doi:10.1016/j.nuclphysa.2022.122447
[arXiv:2103.05419 [physics.ins-det]].
%716 citations counted in INSPIRE as of 18 Jan 2024

%\cite{Strikman:2005yv}
\bibitem{Strikman:2005yv}
M.~Strikman, R.~Vogt and S.~N.~White,
%``Probing small x parton densities in ultraperipheral AA and pA collisions at the LHC,''
Phys. Rev. Lett. \textbf{96}, 082001 (2006)
%doi:10.1103/PhysRevLett.96.082001
[arXiv:hep-ph/0508296 [hep-ph]].
%58 citations counted in INSPIRE as of 17 Jan 2024

%\cite{Baltz:2007kq}
\bibitem{Baltz:2007kq}
A.~J.~Baltz, G.~Baur, D.~d'Enterria, L.~Frankfurt, F.~Gelis, V.~Guzey, K.~Hencken, Y.~Kharlov, M.~Klasen and S.~R.~Klein, \textit{et al.}
%``The Physics of Ultraperipheral Collisions at the LHC,''
Phys. Rept. \textbf{458}, 1-171 (2008)
%doi:10.1016/j.physrep.2007.12.001

%\cite{Guzey:2024gff}
\bibitem{Guzey:2024gff}
V.~Guzey and M.~Strikman,
%``Nuclear suppression of coherent $J/\psi$ photoproduction in heavy-ion UPCs and leading twist nuclear shadowing,''
[arXiv:2404.17476 [hep-ph]].
%0 citations counted in INSPIRE as of 17 May 2024

%\cite{ALICE:2013wjo}
%\bibitem{ALICE:2013wjo}
%E.~Abbas \textit{et al.} [ALICE],
%``Charmonium and $e^+e^-$ pair photoproduction at mid-rapidity in ultra-peripheral Pb-Pb collisions at $\sqrt{s_{\rm NN}}$=2.76 TeV,''
%Eur. Phys. J. C \textbf{73}, no.11, 2617 (2013)
%doi:10.1140/epjc/s10052-013-2617-1
%[arXiv:1305.1467 [nucl-ex]].
%315 citations counted in INSPIRE as of 18 Jan 2024

%\cite{ALICE:2012yye}
%\bibitem{ALICE:2012yye}
%B.~Abelev \textit{et al.} [ALICE],
%``Coherent $J/\psi$ photoproduction in ultra-peripheral Pb-Pb collisions at $\sqrt{s_{NN}} = 2.76$ TeV,''
%Phys. Lett. B \textbf{718}, 1273-1283 (2013)
%doi:10.1016/j.physletb.2012.11.059
%[arXiv:1209.3715 [nucl-ex]].
%311 citations counted in INSPIRE as of 18 Jan 2024

%\cite{CMS:2016itn}
%\bibitem{CMS:2016itn}
%V.~Khachatryan \textit{et al.} [CMS],
%``Coherent $J/\psi$ photoproduction in ultra-peripheral PbPb collisions at $\sqrt {s_{NN}} =$ 2.76 TeV with the CMS experiment,''
%Phys. Lett. B \textbf{772}, 489-511 (2017)
%doi:10.1016/j.physletb.2017.07.001
%[arXiv:1605.06966 [nucl-ex]].
%165 citations counted in INSPIRE as of 18 Jan 2024

%\cite{ALICE:2021gpt}
%\bibitem{ALICE:2021gpt}
%S.~Acharya \textit{et al.} [ALICE],
%``Coherent $J/\psi$ and $\psi'$ photoproduction at midrapidity in ultra-peripheral Pb-Pb collisions at $\sqrt{s_{\mathrm{NN}}}~=~5.02$ TeV,''
%Eur. Phys. J. C \textbf{81}, no.8, 712 (2021)
%doi:10.1140/epjc/s10052-021-09437-6
%[arXiv:2101.04577 [nucl-ex]].
%67 citations counted in INSPIRE as of 18 Jan 2024

%\cite{ALICE:2019tqa}
%\bibitem{ALICE:2019tqa}
%S.~Acharya \textit{et al.} [ALICE],
%``Coherent J/$\psi$ photoproduction at forward rapidity in ultra-peripheral Pb-Pb collisions at $\sqrt{s_{\rm{NN}}}=5.02$ TeV,''
%Phys. Lett. B \textbf{798}, 134926 (2019)
%doi:10.1016/j.physletb.2019.134926
%[arXiv:1904.06272 [nucl-ex]].
%89 citations counted in INSPIRE as of 18 Jan 2024

%\cite{LHCb:2021bfl}
%\bibitem{LHCb:2021bfl}
%R.~Aaij \textit{et al.} [LHCb],
%``Study of coherent $J/\psi$ production in lead-lead collisions at $ \sqrt{{\mathrm{s}}_{\mathrm{NN}}} $ = 5 TeV,''
%JHEP \textbf{07}, 117 (2022)
%doi:10.1007/JHEP07(2022)117
%[arXiv:2107.03223 [hep-ex]].
%29 citations counted in INSPIRE as of 18 Jan 2024

%\cite{LHCb:2022ahs}
%\bibitem{LHCb:2022ahs}
%R.~Aaij \textit{et al.} [LHCb],
%``Study of exclusive photoproduction of charmonium in ultra-peripheral lead-lead collisions,''
%JHEP \textbf{06}, 146 (2023)
%doi:10.1007/JHEP06(2023)146
%[arXiv:2206.08221 [hep-ex]].
%29 citations counted in INSPIRE as of 18 Jan 2024

%\cite{CMS:2023snh}
%\bibitem{CMS:2023snh}
%A.~Tumasyan \textit{et al.} [CMS],
%``Probing Small Bjorken-x Nuclear Gluonic Structure via Coherent J/\ensuremath{\psi} Photoproduction in Ultraperipheral Pb-Pb Collisions at sNN=5.02\,\,TeV,''
%Phys. Rev. Lett. \textbf{131}, no.26, 262301 (2023)
%doi:10.1103/PhysRevLett.131.262301
%[arXiv:2303.16984 [nucl-ex]].
%12 citations counted in INSPIRE as of 18 Jan 2024

%\cite{ALICE:2023jgu}
%\bibitem{ALICE:2023jgu}
%S.~Acharya \textit{et al.} [ALICE],
%``Energy dependence of coherent photonuclear production of J/\ensuremath{\psi} mesons in ultra-peripheral Pb-Pb collisions at $ \sqrt{{\textrm{s}}_{\textrm{NN}}} $ = 5.02 TeV,''
%JHEP \textbf{10}, 119 (2023)
%doi:10.1007/JHEP10(2023)119
%[arXiv:2305.19060 [nucl-ex]].
%10 citations counted in INSPIRE as of 18 Jan 2024

%\cite{STAR:2023gpk}
%\bibitem{STAR:2023gpk}
% [STAR],
%``Exclusive $J/\psi$, $\psi(2s)$, and $e^{+}e^{-}$ pair production in Au$+$Au ultra-peripheral collisions at RHIC,''
%[arXiv:2311.13632 [nucl-ex]].
%2 citations counted in INSPIRE as of 30 Jan 2024

%\cite{Frankfurt:2011cs}
\bibitem{Frankfurt:2011cs}
L.~Frankfurt, V.~Guzey and M.~Strikman,
%``Leading Twist Nuclear Shadowing Phenomena in Hard Processes with Nuclei,''
Phys. Rept. \textbf{512}, 255-393 (2012)
%doi:10.1016/j.physrep.2011.12.002
[arXiv:1106.2091 [hep-ph]].
%206 citations counted in INSPIRE as of 18 Jan 2024

%\cite{Guzey:2013xba}
\bibitem{Guzey:2013xba}
V.~Guzey, E.~Kryshen, M.~Strikman and M.~Zhalov,
%``Evidence for nuclear gluon shadowing from the ALICE measurements of PbPb ultraperipheral exclusive $J/{\psi}$ production,''
Phys. Lett. B \textbf{726}, 290-295 (2013)
%doi:10.1016/j.physletb.2013.08.043
[arXiv:1305.1724 [hep-ph]].
%112 citations counted in INSPIRE as of 18 Jan 2024

%\cite{Guzey:2013qza}
\bibitem{Guzey:2013qza}
V.~Guzey and M.~Zhalov,
%``Exclusive $J/{\psi}$ production in ultraperipheral collisions at the LHC: constrains on the gluon distributions in the proton and nuclei,''
JHEP \textbf{10}, 207 (2013)
%doi:10.1007/JHEP10(2013)207
[arXiv:1307.4526 [hep-ph]].
%98 citations counted in INSPIRE as of 18 Jan 2024

%\cite{Eskola:2022vpi}
\bibitem{Eskola:2022vpi}
K.~J.~Eskola, C.~A.~Flett, V.~Guzey, T.~L\"oyt\"ainen and H.~Paukkunen,
%``Exclusive J/\ensuremath{\psi} photoproduction in ultraperipheral Pb+Pb collisions at the CERN Large Hadron Collider calculated at next-to-leading order perturbative QCD,''
Phys. Rev. C \textbf{106}, no.3, 035202 (2022)
%doi:10.1103/PhysRevC.106.035202
[arXiv:2203.11613 [hep-ph]].
%39 citations counted in INSPIRE as of 30 Jan 2024

%\cite{Eskola:2022vaf}
\bibitem{Eskola:2022vaf}
K.~J.~Eskola, C.~A.~Flett, V.~Guzey, T.~L\"oyt\"ainen and H.~Paukkunen,
%``Next-to-leading order perturbative QCD predictions for exclusive J/\ensuremath{\psi} photoproduction in oxygen-oxygen and lead-lead collisions at energies available at the CERN Large Hadron Collider,''
Phys. Rev. C \textbf{107}, no.4, 044912 (2023)
%doi:10.1103/PhysRevC.107.044912
[arXiv:2210.16048 [hep-ph]].
%15 citations counted in INSPIRE as of 30 Jan 2024


%\cite{Jones:2015nna}
\bibitem{Jones:2015nna}
S.~P.~Jones, A.~D.~Martin, M.~G.~Ryskin and T.~Teubner,
%``Exclusive $J/\psi$ and $\Upsilon$ photoproduction and the low $x$ gluon,''
J. Phys. G \textbf{43}, no.3, 035002 (2016)
%doi:10.1088/0954-3899/43/3/035002
[arXiv:1507.06942 [hep-ph]].
%66 citations counted in INSPIRE as of 01 Feb 2024

%\cite{Jones:2016ldq}
\bibitem{Jones:2016ldq}
S.~P.~Jones, A.~D.~Martin, M.~G.~Ryskin and T.~Teubner,
%``The exclusive $J/\psi$ process at the LHC tamed to probe the low $x$ gluon,''
Eur. Phys. J. C \textbf{76}, no.11, 633 (2016)
%doi:10.1140/epjc/s10052-016-4493-y
[arXiv:1610.02272 [hep-ph]].
%41 citations counted in INSPIRE as of 01 Feb 2024

%\cite{Guzey:2018dlm}
\bibitem{Guzey:2018dlm}
V.~Guzey and M.~Klasen,
%``Inclusive dijet photoproduction in ultraperipheral heavy ion collisions at the CERN Large Hadron Collider in next-to-leading order QCD,''
Phys. Rev. C \textbf{99}, no.6, 065202 (2019)
%doi:10.1103/PhysRevC.99.065202
[arXiv:1811.10236 [hep-ph]].
%23 citations counted in INSPIRE as of 18 Jan 2024

%\cite{Guzey:2016tek}
\bibitem{Guzey:2016tek}
V.~Guzey and M.~Klasen,
%``Diffractive dijet photoproduction in ultraperipheral collisions at the LHC in next-to-leading order QCD,''
JHEP \textbf{04}, 158 (2016)
%doi:10.1007/JHEP04(2016)158
[arXiv:1603.06055 [hep-ph]].

%\cite{ATLAS:2017kwa}
\bibitem{ATLAS:2017kwa}
 [ATLAS],
``Photo-nuclear dijet production in ultra-peripheral Pb+Pb collisions,''
ATLAS-CONF-2017-011.
%43 citations counted in INSPIRE as of 18 Jan 2024

%\cite{ATLAS:2022cbd}
\bibitem{ATLAS:2022cbd}
 [ATLAS],
``Photo-nuclear jet production in ultra-peripheral Pb+Pb collisions at $\sqrt{s}_\text{NN} = 5.02$ TeV with the ATLAS detector,''
ATLAS-CONF-2022-021.
%12 citations counted in INSPIRE as of 18 Jan 2024

%\cite{Pire:2008ea}
\bibitem{Pire:2008ea}
B.~Pire, L.~Szymanowski and J.~Wagner,
%``Can one measure timelike Compton scattering at LHC?,''
Phys. Rev. D \textbf{79}, 014010 (2009)
%doi:10.1103/PhysRevD.79.014010
[arXiv:0811.0321 [hep-ph]].
%67 citations counted in INSPIRE as of 26 Feb 2024

%\cite{Schafer:2010ud}
\bibitem{Schafer:2010ud}
W.~Schafer, G.~Slipek and A.~Szczurek,
%``Exclusive diffractive photoproduction of dileptons by timelike Compton scattering,''
Phys. Lett. B \textbf{688}, 185-191 (2010)
%doi:10.1016/j.physletb.2010.04.009
[arXiv:1003.0610 [hep-ph]].
%15 citations counted in INSPIRE as of 26 Feb 2024

%\cite{Peccini:2021rbt}
\bibitem{Peccini:2021rbt}
G.~M.~Peccini, L.~S.~Moriggi and M.~V.~T.~Machado,
%``Exclusive dilepton production via timelike Compton scattering in heavy ion collisions,''
Phys. Rev. D \textbf{103}, no.5, 054009 (2021)
%doi:10.1103/PhysRevD.103.054009
[arXiv:2101.08338 [hep-ph]].
%7 citations counted in INSPIRE as of 26 Feb 2024

%\cite{Klein:2002wm}
\bibitem{Klein:2002wm}
S.~R.~Klein, J.~Nystrand and R.~Vogt,
%``Heavy quark photoproduction in ultraperipheral heavy ion collisions,''
Phys. Rev. C \textbf{66}, 044906 (2002)
%doi:10.1103/PhysRevC.66.044906
[arXiv:hep-ph/0206220 [hep-ph]].
%63 citations counted in INSPIRE as of 14 Feb 2024

%\cite{Goncalves:2009ey}
\bibitem{Goncalves:2009ey}
V.~P.~Goncalves, M.~V.~T.~Machado and A.~R.~Meneses,
%``Heavy Quark Photoproduction in Coherent Interactions at High Energies,''
Phys. Rev. D \textbf{80}, 034021 (2009)
%doi:10.1103/PhysRevD.80.034021
[arXiv:0905.2067 [hep-ph]].
%15 citations counted in INSPIRE as of 26 Feb 2024

%\cite{Goncalves:2017zdx}
\bibitem{Goncalves:2017zdx}
V.~P.~Gon\c{c}alves, G.~Sampaio dos Santos and C.~R.~Sena,
%``Inclusive heavy quark photoproduction in $pp$, $pPb$ and $PbPb$ collisions at Run 2 LHC energies,''
Nucl. Phys. A \textbf{976}, 33-45 (2018)
%doi:10.1016/j.nuclphysa.2018.05.002
[arXiv:1711.04497 [hep-ph]].
%10 citations counted in INSPIRE as of 26 Feb 2024

\bibitem{Alvioli:2016gfo}
M.~Alvioli, L.~Frankfurt, V.~Guzey, M.~Strikman and M.~Zhalov,
%``Mapping color fluctuations in the photon in ultraperipheral heavy ion collisions at the Large Hadron Collider,''
Phys. Lett. B \textbf{767}, 450-457 (2017)
%doi:10.1016/j.physletb.2017.02.034
[arXiv:1605.06606 [hep-ph]].
%7 citations counted in INSPIRE as of 31 Jan 2024

%\cite{Abramovsky:1973fm}
\bibitem{Abramovsky:1973fm}
V.~A.~Abramovsky, V.~N.~Gribov and O.~V.~Kancheli,
%``Character of Inclusive Spectra and Fluctuations Produced in Inelastic Processes by Multi - Pomeron Exchange,''
Yad. Fiz. \textbf{18}, 595-616 (1973), Sov. J. Nucl. Phys. \text{18},  308-317 (1974).
%855 citations counted in INSPIRE as of 18 Jan 2024

%\cite{Treleani:1994at}
\bibitem{Treleani:1994at}
D.~Treleani,
%``AGK cutting rules and perturbative QCD,''
Int. J. Mod. Phys. A \textbf{11}, 613-654 (1996)
%doi:10.1142/S0217751X96000286
%16 citations counted in INSPIRE as of 01 Feb 2024

%\cite{Jalilian-Marian:2004vhw}
\bibitem{Jalilian-Marian:2004vhw}
J.~Jalilian-Marian and Y.~V.~Kovchegov,
%``Inclusive two-gluon and valence quark-gluon production in DIS and pA,''
Phys. Rev. D \textbf{70}, 114017 (2004)
[erratum: Phys. Rev. D \textbf{71}, 079901 (2005)]
%doi:10.1103/PhysRevD.71.079901
[arXiv:hep-ph/0405266 [hep-ph]].
%191 citations counted in INSPIRE as of 31 Jan 2024

%\cite{Gelis:2006yv}
\bibitem{Gelis:2006yv}
F.~Gelis and R.~Venugopalan,
%``Particle production in field theories coupled to strong external sources,''
Nucl. Phys. A \textbf{776}, 135-171 (2006)
%doi:10.1016/j.nuclphysa.2006.07.020
[arXiv:hep-ph/0601209 [hep-ph]].
%114 citations counted in INSPIRE as of 31 Jan 2024

%\cite{Kovner:2006wr}
\bibitem{Kovner:2006wr}
A.~Kovner and M.~Lublinsky,
%``One gluon, two gluon: Multigluon production via high energy evolution,''
JHEP \textbf{11}, 083 (2006)
%doi:10.1088/1126-6708/2006/11/083
[arXiv:hep-ph/0609227 [hep-ph]].
%83 citations counted in INSPIRE as of 31 Jan 2024

%\cite{Nikolaev:2006mx}
\bibitem{Nikolaev:2006mx}
N.~N.~Nikolaev and W.~Schafer,
%``Unitarity cutting rules for the nucleus excitation and topological cross-sections in hard production off nuclei from nonlinear k-perpendicular factorization,''
Phys. Rev. D \textbf{74}, 074021 (2006)
%doi:10.1103/PhysRevD.74.074021
[arXiv:hep-ph/0607307 [hep-ph]].
%17 citations counted in INSPIRE as of 31 Jan 2024

%\cite{Frankfurt:1991nx}
%\bibitem{Frankfurt:1991nx}
%L.~L.~Frankfurt and M.~I.~Strikman,
%``Diffraction off nuclei in color singlet models of shadowing,''
%Phys. Lett. B \textbf{382}, 6-12 (1996)
%doi:10.1016/0370-2693(96)00653-3
%28 citations counted in INSPIRE as of 19 Jan 2024

%\cite{Frankfurt:1998ym}
\bibitem{Frankfurt:1998ym}
L.~Frankfurt and M.~Strikman,
%``Diffraction at HERA, color opacity and nuclear shadowing,''
Eur. Phys. J. A \textbf{5}, 293-306 (1999)
%doi:10.1007/s100500050288
[arXiv:hep-ph/9812322 [hep-ph]].
%90 citations counted in INSPIRE as of 19 Jan 2024

%\cite{Bertocchi:1976bq}
\bibitem{Bertocchi:1976bq}
L.~Bertocchi and D.~Treleani,
%``Glauber Theory, Unitarity, and the AGK Cancellation,''
J. Phys. G \textbf{3}, 147 (1977)
%doi:10.1088/0305-4616/3/2/007
%78 citations counted in INSPIRE as of 19 Jan 2024%\cite{Alvioli:2016gfo}

%\cite{Bialas:1976ed}
\bibitem{Bialas:1976ed}
A.~Bialas, M.~Bleszynski and W.~Czyz,
%``Multiplicity Distributions in Nucleus-Nucleus Collisions at High-Energies,''
Nucl. Phys. B \textbf{111}, 461-476 (1976)
%doi:10.1016/0550-3213(76)90329-1
%795 citations counted in INSPIRE as of 19 Jan 2024


%\cite{Frankfurt:2022jns}
\bibitem{Frankfurt:2022jns}
L.~Frankfurt, V.~Guzey, A.~Stasto and M.~Strikman,
%``Selected topics in diffraction with protons and nuclei: past, present, and future,''
Rept. Prog. Phys. \textbf{85}, no.12, 126301 (2022)
%doi:10.1088/1361-6633/ac8228
[arXiv:2203.12289 [hep-ph]].
%19 citations counted in INSPIRE as of 31 Jan 2024

%\cite{ParticleDataGroup:2014cgo}
\bibitem{ParticleDataGroup:2014cgo}
K.~A.~Olive \textit{et al.} [Particle Data Group],
%``Review of Particle Physics,''
Chin. Phys. C \textbf{38}, 090001 (2014)
%doi:10.1088/1674-1137/38/9/090001
%9191 citations counted in INSPIRE as of 21 Feb 2024

%\cite{Chapin:1985mf}
\bibitem{Chapin:1985mf}
T.~J.~Chapin, R.~L.~Cool, K.~A.~Goulianos, K.~A.~Jenkins, J.~P.~Silverman, G.~R.~Snow, H.~Sticker, S.~N.~White and Y.~H.~Chou,
%``Diffraction Dissociation of Photons on Hydrogen,''
Phys. Rev. D \textbf{31}, 17-30 (1985)
%doi:10.1103/PhysRevD.31.17
%105 citations counted in INSPIRE as of 21 Feb 2024

%\cite{Alvioli:2014eda}
\bibitem{Alvioli:2014eda}
M.~Alvioli, B.~A.~Cole, L.~Frankfurt, D.~V.~Perepelitsa and M.~Strikman,
%``Evidence for $x$-dependent proton color fluctuations in pA collisions at the CERN Large Hadron Collider,''
Phys. Rev. C \textbf{93}, no.1, 011902 (2016)
%doi:10.1103/PhysRevC.93.011902
[arXiv:1409.7381 [hep-ph]].
%66 citations counted in INSPIRE as of 15 May 2024

%\cite{Alvioli:2017wou}
\bibitem{Alvioli:2017wou}
M.~Alvioli, L.~Frankfurt, D.~Perepelitsa and M.~Strikman,
%``Global analysis of color fluctuation effects in proton\textendash{} and deuteron\textendash{}nucleus collisions at RHIC and the LHC,''
Phys. Rev. D \textbf{98}, no.7, 071502 (2018)
%doi:10.1103/PhysRevD.98.071502
[arXiv:1709.04993 [hep-ph]].
%21 citations counted in INSPIRE as of 15 May 2024

%\cite{Perepelitsa:2024eik}
\bibitem{Perepelitsa:2024eik}
D.~V.~Perepelitsa,
%``Contribution to differential $\pi^0$ and $\gamma_\mathrm{dir}$ modification in small systems from color fluctuation effects,''
[arXiv:2404.17660 [nucl-th]].
%0 citations counted in INSPIRE as of 15 May 2024

%\cite{Alvioli:2009ab}
\bibitem{Alvioli:2009ab}
M.~Alvioli, H.~J.~Drescher and M.~Strikman,
%``A Monte Carlo generator of nucleon configurations in complex nuclei including Nucleon-Nucleon correlations,''
Phys. Lett. B \textbf{680}, 225-230 (2009)
%doi:10.1016/j.physletb.2009.08.067
[arXiv:0905.2670 [nucl-th]].
%86 citations counted in INSPIRE as of 31 Jan 2024

%\cite{Alvioli:2008rw}
\bibitem{Alvioli:2008rw}
M.~Alvioli, C.~Ciofi degli Atti, I.~Marchino, V.~Palli and H.~Morita,
%``Effects of Ground-State Correlations on High Energy Scattering off Nuclei: The Case of the Total Neutron-Nucleus Cross Section,''
Phys. Rev. C \textbf{78}, 031601 (2008)
%doi:10.1103/PhysRevC.78.031601
[arXiv:0807.0873 [nucl-th]].
%24 citations counted in INSPIRE as of 31 Jan 2024

%\cite{Alvioli:2007zz}
\bibitem{Alvioli:2007zz}
M.~Alvioli, C.~Ciofi degli Atti and H.~Morita,
%``Proton-neutron and proton-proton correlations in medium-weight nuclei and the role of the tensor force,''
Phys. Rev. Lett. \textbf{100}, 162503 (2008)
%doi:10.1103/PhysRevLett.100.162503
%103 citations counted in INSPIRE as of 31 Jan 2024

%\cite{ALICE:2021poe}
\bibitem{ALICE:2021poe}
S.~Acharya \textit{et al.} [ALICE],
%``Study of very forward energy and its correlation with particle production at midrapidity in pp and p-Pb collisions at the LHC,''
JHEP \textbf{08}, 086 (2022)
%doi:10.1007/JHEP08(2022)086
[arXiv:2107.10757 [nucl-ex]].
%4 citations counted in INSPIRE as of 22 Jan 2024

%\cite{E665:1995utr}
\bibitem{E665:1995utr}
M.~R.~Adams \textit{et al.} [E665],
%``Nuclear decay following deep inelastic scattering of 470-GeV muons,''
Phys. Rev. Lett. \textbf{74}, 5198-5201 (1995)
[erratum: Phys. Rev. Lett. \textbf{80}, 2020-2021 (1998)]
%doi:10.1103/PhysRevLett.74.5198
%33 citations counted in INSPIRE as of 24 Jan 2024

%\cite{Strikman:1998cc}
\bibitem{Strikman:1998cc}
M.~Strikman, M.~G.~Tverskoii and M.~B.~Zhalov,
%``Soft neutron production in DIS: A Window to the final state interactions,''
Phys. Lett. B \textbf{459}, 37-42 (1999)
%doi:10.1016/S0370-2693(99)00627-9
[arXiv:nucl-th/9806099 [nucl-th]].
%20 citations counted in INSPIRE as of 24 Jan 2024

%\cite{Larionov:2018igy}
\bibitem{Larionov:2018igy}
A.~B.~Larionov and M.~Strikman,
%``Slow neutron production as a probe of hadron formation in high-energy $\gamma^*A$ reactions,''
Phys. Rev. C \textbf{101}, no.1, 014617 (2020)
%doi:10.1103/PhysRevC.101.014617
[arXiv:1812.08231 [hep-ph]].
%6 citations counted in INSPIRE as of 22 Jan 2024

%\cite{Alvioli:2010yk}
\bibitem{Alvioli:2010yk}
M.~Alvioli and M.~Strikman,
%``Beam Fragmentation in Heavy Ion Collisions with Realistically Correlated Nuclear Configurations,''
Phys. Rev. C \textbf{83}, 044905 (2011)
%doi:10.1103/PhysRevC.83.044905
[arXiv:1008.2328 [nucl-th]].
%20 citations counted in INSPIRE as of 19 Feb 2024

%\cite{Chang:2022hkt}
\bibitem{Chang:2022hkt}
W.~Chang, E.~C.~Aschenauer, M.~D.~Baker, A.~Jentsch, J.~H.~Lee, Z.~Tu, Z.~Yin and L.~Zheng,
%``Benchmark eA generator for leptoproduction in high-energy lepton-nucleus collisions,''
Phys. Rev. D \textbf{106}, no.1, 012007 (2022)
%doi:10.1103/PhysRevD.106.012007
[arXiv:2204.11998 [physics.comp-ph]].
%14 citations counted in INSPIRE as of 24 Jan 2024

%\cite{ATLAS:2015hkr}
\bibitem{ATLAS:2015hkr}
G.~Aad \textit{et al.} [ATLAS],
%``Measurement of the centrality dependence of the charged-particle pseudorapidity distribution in proton\textendash{}lead collisions at $\sqrt{s_{_\text {NN}}} = 5.02$  TeV with the ATLAS detector,''
Eur. Phys. J. C \textbf{76}, no.4, 199 (2016)
%doi:10.1140/epjc/s10052-016-4002-3
[arXiv:1508.00848 [hep-ex]].
%99 citations counted in INSPIRE as of 26 Feb 2024

%\cite{Alvioli:2013vk}
%\bibitem{Alvioli:2013vk}
%M.~Alvioli and M.~Strikman,
%``Color fluctuation effects in proton-nucleus collisions,''
%Phys. Lett. B \textbf{722}, 347-354 (2013)
%doi:10.1016/j.physletb.2013.04.042
%[arXiv:1301.0728 [hep-ph]].
%90 citations counted in INSPIRE as of 31 Jan 2024

%\cite{Alvioli:2014sba}
\bibitem{Alvioli:2014sba}
M.~Alvioli, L.~Frankfurt, V.~Guzey and M.~Strikman,
%``Revealing \textquotedblleft{}flickering\textquotedblright{} of the interaction strength in pA collisions at the CERN LHC,''
Phys. Rev. C \textbf{90}, 034914 (2014)
%doi:10.1103/PhysRevC.90.034914
[arXiv:1402.2868 [hep-ph]].
%31 citations counted in INSPIRE as of 31 Jan 2024

%\cite{Alvioli:2019kcy}
\bibitem{Alvioli:2019kcy}
M.~Alvioli, M.~Azarkin, B.~Blok and M.~Strikman,
%``Revealing minijet dynamics via centrality dependence of double parton interactions in proton\textendash{}nucleus collisions,''
Eur. Phys. J. C \textbf{79}, no.6, 482 (2019)
%doi:10.1140/epjc/s10052-019-6998-7
[arXiv:1901.11266 [hep-ph]].
%10 citations counted in INSPIRE as of 14 Feb 2024

%\cite{ATLAS:2023zfx}
%\bibitem{ATLAS:2023zfx}
%G.~Aad \textit{et al.} [ATLAS],
%``Measurement of the centrality dependence of the dijet yield in $p$+Pb collisions at $\sqrt{s_{_\text{NN}}}$ = 8.16 TeV with the ATLAS detector,''
%[arXiv:2309.00033 [nucl-ex]].
%0 citations counted in INSPIRE as of 31 Jan 2024

\end{thebibliography}
 \end{document}